# TRANSITION REGION EMISSION FROM SOLAR FLARES DURING THE IMPULSIVE PHASE

H. JOHNSON[1,2], J.C. RAYMOND[1], N.A. MURPHY[1], S. GIORDANO[3], Y.-K. KO[4], A. CIARAVELLA[5] & R. SULEIMAN[1]
*Draft version April 28, 2011*

## ABSTRACT

There are relatively few observations of UV emission during the impulsive phases of solar flares, so the nature of that emission is poorly known. Photons produced by solar flares can resonantly scatter off atoms and ions in the corona. Based on off-limb measurements by SOHO/UVCS, we derive the O VI $\lambda1032$ luminosities for 29 flares during the impulsive phase and the Ly$\alpha$ luminosities of 5 flares, and we compare them with X-ray luminosities from GOES measurements. The upper transition region and lower transition region luminosities of the events observed are comparable. They are also comparable to the luminosity of the X-ray emitting gas at the beginning of the flare, but after 10-15 minutes the X-ray luminosity usually dominates. In some cases we can use Doppler dimming to estimate flow speeds of the O VI emitting gas, and 5 events show speeds in the 40 to 80 km s$^{-1}$ range. The O VI emission could originate in gas evaporating to fill the X-ray flare loops, in heated chromospheric gas at the footpoints, or in heated prominence material in the coronal mass ejection. All three sources may contribute in different events or even in a single event, and the relative timing of UV and X-ray brightness peaks, the flow speeds, and the total O VI luminosity favor each source in one or more events.

*Subject headings:*

## 1. INTRODUCTION

Measurements of the energy budget of solar flares are a basic requirement for understanding the physical processes in solar eruptions. Such measurements are difficult, however, because they require simultaneous observations in many wavelength bands. The energy budget during the impulsive phase is especially important, but it is also especially difficult to obtain because of the rapid transformations among energy of nonthermal particles, kinetic energy and thermal energy. Energy budgets for a few events are given in Holman et al. (2003), Emslie et al. (2005) and Raymond et al. (2007), but the largest term is likely to be the white light continuum, and that is poorly determined (Fletcher et al. 2007). Over the lifetime of the flare, the kinetic energy of the associated coronal mass ejection (CME) and the thermal energy deposited in the CME plasma are comparable to the radiation from the X-ray emitting plasma (Emslie et al. 2005; Akmal et al. 2001).

In the standard picture of solar eruptions, the impulsive phase is a brief (10 to 30 minutes) period when reconnection produces energetic particles in the corona. In most large events a CME lifts off at about the same time. Some of the energetic particles stream down to the chromosphere, where they produce hard X-rays and heat the plasma to temperatures around $10^7$ K. The heated gas flows up to fill the flare loops, producing the X-ray emission during the gradual phase. Slit spectrometers seldom catch more than a glimpse of the impulsive phase, so UV spectra are rarely available. Therefore, the energy budget of the evaporating gas at transition region temperatures is particularly difficult to determine. Full sun measurements with SOLSTICE and EVE (Brekke et al. 1996;

Woods et al., 2005; Martínez Oliveros et al. 2010) show that these instruments can obtain the luminosities at transition region temperatures in bright flares, but few measurements are available.

Raymond et al. (2007) presented a technique to measure flare UV photons that are scattered in the corona. In particular, the Ultraviolet Coronagraph Spectrometer (UVCS) aboard SOHO measured O VI $\lambda1032$ photons from flares that are resonantly scattered by O VI ions in the corona. Most flares are associated with coronal mass ejections (CMEs). When the CME reaches the position of the UVCS slit, it blows the pre-event corona away, and the intensity of scattered photons rapidly declines. Thus this technique provides intensities only for 5 to 15 minutes in the impulsive phase of the flare. In favorable cases, it is also possible to estimate the flow speed of the O VI emitting plasma from differential Doppler dimming along the UVCS slit. Raymond et al. (2007) determined the O VI and transition region luminosities and compared them with X-ray temperatures and luminosities from GOES and RHESSI data for five events. Here we expand the sample to 29 events and add a similar analysis for flare Ly$\alpha$ emission in five events. A few events occurred behind the solar limb, which affects the X-ray observations but not the UV intensity.

In the four events for which we have both Ly$\alpha$ and O VI luminosities, the Ly$\alpha$ luminosity is a few times larger than the O VI luminosity and roughly comparable to the luminosity of the upper transition region. Overall the total emission from the transition region is comparable to the emission from the thermal X-ray emitting gas during the early impulsive phase, but radiation from the X-ray emitting loops dominates by the end of the impulsive phase.

There are three likely sources for the O VI emission. In the standard flare picture, chromospheric plasma is heated to $\sim 10^7$ K and fills the flare X-ray loops. In the process, it must be ionized through the O VI ionization state and produce some $\lambda1032$ emission. The standard flare picture also predicts that much of the flare energy is deposited in the chromosphere by energetic particles. Most of that energy is radiated away in the white light continuum (Fletcher et al. 2007), but some may emerge in UV emission lines. Finally, the prominence material in CMEs is strongly heated (Filippov & Koutchmy 2002; Akmal et al. 2001), and bright emission from transition region temperatures has been reported (Ciaravella et al. 2000; Landi et al. 2010). We consider the correlations between X-ray and UV intensities expected in these three cases and compare to the observations.

[1] Harvard-Smithsonian Center for Astrophysics, 60 Garden Street, Cambridge, MA 02138

[2] Northeastern University Physics Department, 110 Forsyth St., 111 Dana Research Center, Boston, MA 02115

[3] INAF-Osservatorio Astronomico di Torino, via Osservatorio 20, 10025 Pino Torinese, Italy

[4] Space Science Division, Naval Research Laboratory, Washington, DC 20375

[5] INAF-Osservatorio Astronomico di Palermo, P.za Parlamento 1, 90134 Palermo, Italy

The observations are presented in §2, and §3 gives the derivation of O VI λ1032 and transition region luminosities. §4 presents the results and compares UV and X-ray emission, and §5 summarizes the paper. Details about some individual events are presented in the Appendix.

## 2. OBSERVATIONS

### 2.1. *UVCS*

The Ultraviolet Coronagraph Spectrometer (UVCS; Kohl et al. 1995, 1997) on SOHO performs ultraviolet spectroscopy in the extended solar corona between 1.5 and 10 solar radii. It has both an O VI channel and a Lyman α channel, and it also measures lines such as Si XII, C III and Mg X. In this report, we will concentrate mainly on the O VI spectral line doublet (at 1032 and 1037 Å) while including some Lyman α observations.

Intensities for the O VI λλ1032, 1037 lines were measured using the Data Analysis Software (DAS40) for UVCS. This version includes the latest radiometric calibration based on repeated observations of stars (Gardner et al. 2002). The 42′ slit was divided into 5 or 6 portions and the profile of each portion was fitted to a Gaussian.

We observed a total of 30 events, 3 of which occurred behind the limb, and 5 of which were covered in the previous paper by Raymond et al. (2007); we reanalyze the events from the previous paper for the purpose of consistency. The events were found during a search of the CDAW LASCO CME catalog for events where the UVCS slit was positioned roughly over the flare. The resulting catalog of UVCS CME observations includes over 800 events. The 30 events presented here show the characteristics of flare radiation resonantly scattered by coronal O VI ions or H I atoms; the emission brightens simultaneously along the entire UVCS slit, it precedes any other CME emission, and it shows no Doppler shift or increased line width compared to the pre-event line profile. A key requirement is that the intensity ratio of O VI $I(1032)/I(1037)$ be consistent with 4:1, indicating scattering of disk radiation. Collisional excitation in the corona produces a 2:1 ratio. In some cases the emission is faint near the ends of the slit, and sometimes the signal-to-noise ratio of the O VI intensity ratio is poor, but we are confident that all the events presented are examples of flare radiation scattered in the corona.

The observation parameters for each event are shown in Table 1. We use the IAU flare naming convention and give the GOES X-ray class, the flare position, the height and position angle of the UVCS slit, the slit width and the exposure time for each event. The flare classes for some events are not available because X-ray flux from larger flares occurring on other parts of the disk at about the same time overwhelmed the X-rays from the relevant smaller flare. Further information about the UVCS observations is available in the pages of the UVCS CME catalog, which are linked to the events listed in the CDAW LASCO CME catalog (http://cdaw.gsfc.nasa.gov/CME_list/). Detailed analyses of the UVCS observations of several events have been presented by Raymond et al. (2003) and Raymond et al. (2007). UVCS could only detect flares close to the limb, and it missed some events either because the operating program at the time was unfavorable, for reasons such as changes in slit height or because the coronal density was too small.

### 2.2. *GOES*

X-ray observations from the GOES satellite were used to obtain thermal and X-ray emission measure values for each event. To do this, we used the Solarsoft Package for GOES written by Stephen White. The ratio between hard and soft X-ray bands was used to find the temperature, and the flux in the soft X-ray band was used to find the emission measure. In most cases we have observations from two GOES satellites and use the average of the two temperatures. The temperatures are consistent except for the 1999 Dec 28 event, in which the temperatures differed by 10 MK. For each exposure, we use the emission measure, temperature and total emissivity to find the thermal luminosity of the X-ray emitting plasma. There may be a nonthermal contribution to the X-ray flux, in which case we overestimate the flare temperature and X-ray luminosity. In three cases the flare was behind the limb (1998 May 11, 1999 Dec. 26 and 2001 Aug 11 17:26 UT), and in the case of the 2003 November 4 flare, the GOES detectors were saturated beginning at 19:42 UT. The flare on 2001 Dec 28 was obscured by a much brighter flare that occurred about 1.5 hours earlier.

Figures 1 - 4 show curves for the O VI 1032 intensity integrated over the entire UVCS slit and the GOES hard and soft X-ray fluxes for each event. The pre-event O VI intensity is subtracted, so that the intensities start near zero. Some events show influence from the CME. In many cases the CME removes O VI from the line of sight, so that the O VI brightens for several exposures due to illumination by the flare, then drops below zero when the CME front reaches the slit and removes O VI ions from the UVCS field of view. Subsequent O VI brightenings sometimes occur when relatively cool, dense CME plasma reaches the UVCS slit.

## 3. ANALYSIS

Deriving the transition region luminosity during the impulsive phase requires a number of steps. First, we observe the pre-flare corona to determine the column density of O VI. From this, we use the scattering cross section to find the illuminating flux during each flare. Last, we use the ratio of O VI luminosity to the transition region luminosity calculated from cooling rates (Raymond et al. 2007) to derive the transition region luminosity. Doppler dimming can cause the derived illuminating flux to vary along the slit (Raymond et al. 2007), and we use this to determine the upflow speed and Doppler dimming correction.

### 3.1. *Preflare Calculations.*

To find the O VI column density, we must first look at the flux ratio of O VI(1032)/O VI(1037), which is typically around 3:1 in the quiet corona. These intensities originate from both collisional excitation and scattered O VI photons. The collisional component gives an intensity ratio $I_{coll}(1032)/I_{coll}(1037)$ of 2:1, while the radiative component $I_{rad}(1032)/I_{rad}(1037)$ gives a ratio of 4:1. From the observed intensity ratio, we can find the ratio of $I_{coll}(1032):I_{rad}(1032)$. With this value, we can find the electron density, given that

$$\frac{I_{coll}(1032)}{I_{rad}(1032)} = \frac{q_{1032}n_e n_{OVI}}{n_{OVI}\sigma_{eff}WI_{disk}(1032)} \tag{1}$$

where $q_{1032}$ is the collision excitation rate from the CHIANTI atomic database (Dere et al., 1997, 2009), $n_e$ and $n_{OVI}$ are the electron and O VI densities, $\sigma_{eff}$ is the effective scattering cross section, $I_{disk}$ is the intensity of the line at the solar surface and $W = 2\pi(1 - \sqrt{1 - 1/r^2})$ is the dilution factor averaged along the line of sight (where r is the heliocentric height). We use a value of $1.62 \times 10^{-8}$ cm$^3$/s for $q_{1032}$ in all events, as the excitation rate varies only weakly at coronal temperatures. The value for $\sigma_{eff}$ is found by convolving the scattering profile and O VI emission profile, and it equals

$$\sigma_{eff} = \frac{\pi e^2}{m_e c} f_{line} \frac{\lambda}{\delta v},$$ (2)

where $f_{line}$ is the oscillator strength (0.131 for O VI $\lambda$ 1032) and $\delta v$ is the assumed coronal line width, here 60 km s$^{-1}$ (Kohl et al., 1995, 1997). $W$ is the dilution factor, which incorporates the solid angle subtended by the entire disk. We assume that the scattering plasma is located near the plane of the sky, because active region streamers generally overlie the flare sites. The disk intensities are taken from UARS and SORCE irradiance measurements (Woods et al., 2004, 2005).

Next, the O VI column density, $N_{OVI}$, can be derived using $n_e$ from equation (1).

$$N_{OVI} = \frac{4\pi I_{coll}(1032)}{q_{1032} n_e} \text{cm}^{-2}.$$ (3)

### 3.2. Flare Calculations

Using the value of the O VI column density found from the pre-flare calculations, we can derive the luminosity for each exposure of the flare given the relationship

$$L_{flare}(1032) = \frac{4\pi I_{flare}(1032)}{N_{OVI} \sigma'_{eff} W'} \text{ photons s}^{-1}$$ (4)

where $I_{flare}$ is the O VI intensity observed by UVCS above the pre-event intensity. Here, $W' = 1/4\pi r'^2$ , where $r'$ is the distance from the flare to a point on the UVCS slit. This differs from the dilution factor, W, in the pre-flare calculations because the luminosity originates from essentially a point source rather than the entire disk. Flare positions are listed in Table 1. For events behind the limb, we assumed that the flare occurred below the UVCS reference point.

The cross-section $\sigma'_{eff}$ in equation (4) differs from $\sigma_{eff}$ in the pre-flare calculations in that it can account for changes due to Doppler dimming effects. In the evaporation picture, the upflow should be directed approximately at the position on the UVCS slit that lies directly over the flare, while the Doppler shift seen at other points along the slit is reduced by the cosine of the angle from the vertical. Therefore, we can use the requirement that the derived luminosity must be the same for each portion of the slit to constrain the upflow velocity using Doppler dimming. Doppler dimming values can be found from the equation

$$DD = \frac{\int e^{-(\frac{v}{\delta v_{TR}})^2} e^{-(\frac{v - v_{dopp}}{\delta v_{cor}})^2}}{\int e^{-(\frac{v}{\delta v_{TR}})^2} e^{-(\frac{v}{\delta v_{cor}})^2}}$$ (5)

assuming a coronal line width, $\delta v_{cor}$, of 60 km s$^{-1}$ and a disk emission profile width, $\delta v_{TR}$, of 30 km s$^{-1}$. After evaluating the integral with these assumptions, we find that the Doppler dimming value can be simplified to

$$DD = e^{-(\frac{v_{dopp}}{67})^2}$$ (6)

For our study, we tested speeds up to 100 km s$^{-1}$ for each event and computed the angle between the vertical and the position on the UVCS slit based on the flare positions given in Table 1. When we multiply $\sigma'_{eff}$ by the Doppler dimming constant for each position along the slit, we start to see a correction to the apparent luminosity. While the uncorrected values show a dip directly above the flare site, the properly corrected values must be constant along the slit. We can estimate the upflow speed by observing which value creates the most uniform derived luminosity.

Figure 5 shows one exposure of the November 04, 2003 event, with luminosities corrected for outflow speeds ranging from 10-100 km s$^{-1}$. At speeds below 80 km s$^{-1}$, we see a dip in the luminosity, while above we see a small peak. From this, we can suppose that the outflow velocity is around 80 km s$^{-1}$. This is close to the estimate of 90 km s$^{-1}$ found in Raymond et al. (2007). Other events show similar characteristics, but not all. The events that show outflow speeds are noted in Table 2. In many cases the signal-to-noise ratio in the individual bins was too poor to permit a flow speed to be determined, especially if the coronal density near the ends of the slit was low. When the flow speed is measured, it is probably accurate to about 20 km s$^{-1}$.

Table 2 presents values for the O VI $\lambda$1032 luminosity, transition region luminosity, X-ray luminosity, X-ray and ultraviolet emission measures, and X-ray temperature for each exposure for all events. The time shown is the beginning of the UVCS exposure. The last one or two exposures in each event might be affected by the CME, so they could underestimate the O VI and transition region fluxes.

In the five cases listed in Table 3 we were able to carry out a similar analysis for Ly$\alpha$. The analysis is somewhat simpler, because the pre-event Ly$\alpha$ originates almost entirely from resonance scattering (Raymond et al. 1997), and the line width of about 250 km s$^{-1}$ is large enough to make Doppler dimming unimportant. One event was observed with the LYA channel of UVCS. For the 1998 January 20 event, the slit width was 50$\mu$m, and the other parameters are the same as those given in Table 1. One event in Tables 1 and 3, 2005 July 9, is not included in Table 2, as the O VI lines were not observed.

We estimate the uncertainties in the derived O VI and Ly$\alpha$ luminosities to be less than a factor of 2. This incorporates the uncertainty from line width, the Doppler dimming estimates considering the signal to noise ratio, and the ratio of O VI $\lambda$1032 luminosity to the transition region luminosity, which depends on abundances and ionization states. In addition, we use the assumption that the O VI is concentrated along the line of sight near the plane of the sky in streamers located above the flare active regions. If the streamer is far from the plane of the sky, the coronal gas is farther from the flare site, and we underestimate the O VI luminosity.

## 4. DISCUSSION

In the standard picture of solar flares we expect that O VI emission during the impulsive phase arises from gas that is being rapidly heated. It could be evaporating gas that is being heated toward the 10 MK temperatures of the post-flare loops, it might be chromospheric plasma heated by energetic particles or shocks in the footpoints, in which case either upflows or downflows might be seen, or it could be plasma in the ejected prominence that is heated as it is being accelerated. We consider these three likely emission sites in turn.

### 4.1. Origin of the Transition Region emission

*Chromospheric Evaporation:* If evaporation of chromospheric gas fills the hot loops that emit X-rays, upflows should be present in all cases. For typical loop lengths and impulsive phase durations, the required speed is tens of km s$^{-1}$. The thermal X-ray emission should peak when the evaporation is nearly complete, so the peak O VI emission should precede the peak in soft X-rays. The X-ray emission measure at a given time is proportional to the amount of gas evaporated at earlier times. Therefore, one expects that the X-ray emission is roughly proportional to the time integral of the transition region emission up until that time. For an evaporated mass $m$, the proportionality is determined by temperature and density of the X-ray emitting plasma, with the emission measure proportional to $mn_e$, and by the O VI emission per unit mass during the evaporation, which is proportional to the excitation rate times the time spent in the O VI ionization state. Thus there should be some correlation between $L_X$ and the integral of $L_{TR}$, but with scatter due to differences among the flares in density and heating rate.

*Footpoint Emission:* If energetic particles deposit energy in the chromosphere and the plasma radiates that energy away as white light continuum or UV emission lines, the UV emission should be roughly proportional to the brightness of nonthermal X-rays, but only indirectly related to the thermal X-ray luminosity. If evaporation occurs at the same time as intense chromospheric heating, the timing of UV and X-ray peaks should be similar to that expected in the chromospheric evaporation scenario. Expansion of the heated gas can produce either upflows or downflows at speeds of a few km s$^{-1}$. The density in the UV emitting gas will be very high, but probably not high enough to thermalize the emission and affect the 2:1 intensity ratio of the $\lambda\lambda1032, 1037$ lines ($n_{crit} \sim 10^{16}$ cm$^{-3}$). The high density implies a cooling time on the order of a second, so large variations in $L_X/L_{TR}$ may occur during an event.

*CME emission:* Erupting prominences often change from absorption to emission in EUV images such as EIT 195 Å as they rise a few tenths of a solar radius above the surface (Filippov & Koutchmy 2002), indicating a temperature increase from $10^4$ K to $2 \times 10^5$ K where O V is produced (Ciaravella et al. 2000) or $> 10^6$ K where the Fe XII lines that normally dominate that band are formed. The heating continues as the gas moves to larger heights (Lee et al. 2009), but expansion rapidly reduces the emission even if the ionic fractions of transition region ions remain high. There has been little systematic study of transition region emission from erupting prominences during the early phases, but there seems to be enormous variation among events. Huge plumes of He II $\lambda304$ emission are sometimes seen, while some events show eruptions only in the highest temperature bands (Reeves & Golub 2011). Landi et al. (2010) detected gas at transition region temperatures during a modest flare at 0.15 R$_\odot$ above the surface with a speed of about 90 km s$^{-1}$ and an emission measure of a few times $10^{45}$ cm$^{-3}$, or about a percent of the values of $EM_{TR}$ shown in Table 2. Heating of the erupting plasma can most easily be attributed to dissipation of magnetic free energy (Kumar & Rust 1996) or injection of kinetic, thermal or energetic particle energy from the reconnection current sheet (Landi et al. 2010; Murphy et al. 2011; Reeves et al. 2010).

The properties of the emission from the erupting promi-nence are very hard to predict. The mass involved is large, but the density declines rapidly. The heating might be expected to accompany CME acceleration, which generally correlates with hard X-rays and the derivative of the thermal X-ray emission (Zhang et al. 2004), though CME initiation can precede or follow the X-ray onset. Since the amount of prominence material varies a great deal among different events, the correlation between $L_{TR}$ and $L_X$ is likely to be loose.

### 4.2. Comparison of expectations with observations

Both upflows and downflows of tens of km s$^{-1}$ are observed in UV lines, and larger speeds are sometimes seen in X-ray lines. (Doschek, Mariska & Sakao 1996; Innes 2001; Milligan et al. 2006; Li & Ding 2010; Veronig et al. 2010). In the events where we can infer upflow speeds, we find 40-80 km s$^{-1}$ in all five cases. Those values are somewhat high, but not anomalous, compared to direct spectroscopic observations at transition region temperatures. However, it should be noted that our method is not sensitive to lower upflow speeds. At upflow speeds above 100 km s$^{-1}$ the O VI emission would be severely Doppler dimmed, and the 4:1 ratio of the O VI doublet would be reduced, so that we would probably not detect events with higher upflow speeds. For the 5 events in which we see upflows, either an evaporation or a CME origin would be favored.

To test for a correlation between X-ray emission and the amount of evaporated material expected in the evaporation picture, we show the change in $L_X$ between the second and third UVCS exposures plotted against the total transition region emission during the second exposure, $L_{TR}$ times the exposure time, in Figure 6. We chose the second and third exposures because the first exposure often has lower a signal-to-noise ratio, and because 8 of the events have no fourth exposure. Events behind the limb are excluded, as are six events in which $L_X$ decreases between exposures 2 and 3. The dashed line is a least squares fit with a slope of 1.76. The scatter is considerable, but the correlation is clear. We conclude that in the simple picture where the transition region emission arises from plasma evaporating to fill the X-ray loops, factors such as the pressure or the evaporation speed must vary considerably from event to event. The fact that $L_X$ decreased in six events might be explained by cooling or expansion on timescales comparable to the UVCS exposure times. However, the fact that 6 of the 29 events show a drop in $L_X$ while $L_{TR}$ is still high is more easily compatible with the footpoint or CME pictures.

Figure 7 shows another comparison between X-rays and transition region emission. This time $L_X$ is plotted against $L_{TR}$ at the peak of the $L_{TR}$. This plot also shows a strong correlation with substantial scatter. Note that the events with large $L_{TR}$ and small $L_X$ are behind the limb. By the time of the peak, a large amount of hot gas has accumulated in the loops, and $L_X$ is about 10 $L_{TR}$. If one requires a fit that goes through the origin and excludes events behind the limb, it would be much steeper than linear; $L_X \propto L_{TR}^{1.54}$. That suggests that the X-ray emitting gas in more energetic events accumulates in higher pressure loops than in less energetic events. In some of the most energetic X-ray events there is little or no cool gas seen in the CMEs (e.g., 4 November 2003, Raymond et al. 2007 and 23 July 2002, Raymond et al. 2003), so the correlation in Figure 7 tends to disfavor an erupting prominence origin for the O VI emission. The loose correlation is compatible with evaporation or footpoint origins.

In the evaporation scenario, one also expects that the thermal X-ray emission peaks after the emission from the evaporation flow as the loops fill with hot plasma. The relationship between the X-ray luminosity and the transition region luminosity can be seen in Figure 8, a plot of the GOES peak flux against the time lag from the O VI peak to the GOES soft X-ray peak. Either the two are unrelated (28 Dec. 2001), the GOES peaks at roughly the same time as the O VI intensity (e.g., 28 Jul. 2004, 6 Feb. 2000), or the GOES peak occurs later than the O VI peak (e.g., 23 Oct. 2004, 4 Nov. 2003). Larger flares will usually have the GOES peak later than the O VI peak, while in smaller flares it can either lag or be simultaneous. Unrelated GOES events could point to events that occur behind the limb, but not always. The CME corresponding to the flare may arrive at the slit before the X-ray flux peaks, removing the scattering O VI ions. Figure 8 only includes events in which the lag is positive. Again there is a correlation, but with large scatter. The plot is consistent with the idea that flares develop over time, with larger flares taking longer to energize the large number of the magnetic loops that become part of the flare. The large number of events with a negative time lag is unlikely in the evaporation picture, and unexpected in the footpoint picture, so it tends to favor O VI emission from the erupting prominence.

We can also consider the time scale of the heating. If plasma is instantaneously heated from a low temperature to a flare temperature of $\sim 10^7$ K, each oxygen atom spends a time $t_i = 1/(n_e q_{ion})$ given by the ionization rate $q_{ion}$ as O VI, and it is excited at a rate $n_e q_{ex}$. It will emit a number of photons equal to $q_{ex}/q_{ion}$ before being ionized, or about 11 photons per O atom. Thus it would be necessary to heat about $\dot{m} = 2 \times 10^{14}$ g s$^{-1}$ to produce a $\lambda 1032$ luminosity of $10^{25}$ erg s$^{-1}$. On the other hand, the mass of the X-ray emitting gas can be estimated from the emission measure and a typical flare density of $10^{11}$ cm$^{-3}$ (Del Zanna et al. 2011; Reznikova et al. 2009), m $= 2 \times 10^{13} EM_{48}/n_{11}$ g. Since the O VI emission is observed for several hundred seconds, $\dot{m}$ must be less than $10^{11}$ g s$^{-1}$. We conclude that if the O VI luminosity comes mostly from evaporating gas, it is produced not by gas that is heated on a time scale $t_i \sim 0.01$ sec, but by gas that is heated on a time scale on the order of tens of seconds. Thus in the evaporation picture, the O VI emission must originate as the gas is gradually heated through transition region temperatures. The mass of an erupting prominence can be a significant fraction of the CME mass of $10^{14}$ to $10^{16}$ g (Vourlidas et al. 2010), so more rapid heating would be possible or even necessary if the O VI originates in the CME.

We can make a few statements regarding the energy budget. The values of $L_{Ly\alpha}$ are comparable to the values of $L_{TR}$ in the four cases where we can make the comparison, but the scatter is large. Overall, it seems that the emission from the lower transition region is comparable to that of the upper transition region. Emission in the He II $\lambda 304$ line can be substantial (e.g. Martínez Oliveros 2011), quite likely on the order of the Ly$\alpha$ emission. Altogether, the transition region luminosity is typically comparable to $L_X$ at the beginning of an event, but the luminosity of the X-ray emitting gas increases to several times the transition region emission after roughly 10 minutes. In any case, white light continuum from the chromosphere is likely to be 100 times larger (Fletcher et al. 2007), dominating the total energy budget of the flare.

## 5. SUMMARY

Transition region luminosities were measured for the impulsive phases of 30 solar flares. The origin of the emission could be chromospheric evaporation, footpoint emission or heated prominence ejecta. None of the three is strongly favored, but the evaporation seems most likely in the events where the UV peak precedes the X-ray peak. Five events show 40 to 80 km s$^{-1}$ velocity shifts, suggesting an evaporation or erupting prominence origin. A combination of emission sites is of course possible. The models of Allred et al., (2005) show a phase of gentle heating, in which radiative losses balance the energy deposition, followed by an explosive phase when heating dominates and strong flows occur, so emission from the footpoints might explain the behavior seen in the 2002 July 23 event.

The upper and lower transition region luminosities are similar in the four events where the comparison can be made. They are comparable to the emission from the thermal X-ray emitting gas observed by GOES early in the events, but radiative cooling of the X-ray emitting gas begins to dominate by the end of the impulsive phase. This indicates that thermal conduction does not carry the internal energy of the flare loops to be radiated away at transition region temperatures. It is unlikely that it passes through the transition region to be radiated away at lower temperatures because of the extremely steep temperature gradients and relatively low pressures that would require, but it is still possible that conductive heating drives the evaporative flow.

Our study is limited to two bright lines and to the first 10-15 minutes of the flare. Future analyses of UV and EUV emission lines with the Extreme ultraviolet Variability Experiment (EVE), should help to determine the origin of the emission and pin down its contribution to the energy budget.

This work was supported by NASA grants NNG06GG78G and NNX09AB17G-R to the Smithsonian Astrophysical Observatory. SOHO is a project of international cooperation between ESA and NASA.

*Facilities:* SOHO (UVCS)

## APPENDIX

Here we some comment on some individual events.

*1998 May 11:* This event occurred behind the solar limb, so much of the X-ray emission was obscured. From the O VI luminosity and the correlation in Figure 6, it seems likely that this would have been very roughly an X8 event had it occurred on the near side of the Sun.

*1999 Dec 26:* This event occurred behind the solar limb, so much of the X-ray emission was obscured. From the O VI luminosity and the correlation in Figure 6, it seems likely that this would have been very roughly an X10 event had it occurred on the near side of the Sun.

*1999 Dec 28:* Torsti et al. (2002) analyzed radio, hard X-ray, gamma ray and SEP observations of this event. The O VI emission seen by UVCS corresponds closely in time to the hard X-ray emission and an intense type III radio burst. An SEP event with enhanced He and $^3$He was seen by ERNE, with an inferred release time of 01:20 UT.

*2000 Feb 05:* This event was far from the solar limb compared to most of the other flares in this study. There were no EIT observations. It produced a strong type III burst (Reiner et al. 2001), and it may have generated a shock that produced kilometric type II emission on 7 February.

*2000 Aug 06:* There were no EIT observations of this event.

*2001 Aug 11:* The second event on Aug. 11 occurred behind the limb.

*2001 Dec 28:* The X-ray emission from this event was overshadowed by a much brighter event that occurred about 1.5 hours earlier.

*2002 Jul 23:* This flare has been extensively studied at wavelengths from radio to gamma rays. UVCS observations of this event are analyzed in Raymond et al. (2003) and Raymond et al. (2007), and the energy budget is presented by Emslie et al. (2005). RHESSI X-ray observations show a rise phase beginning at 00:18 UT and a sudden hard X-ray increase beginning at 00:27. The hard X-rays and gamma rays peaked between 00:26 and 00:28 (Lin et al. 2003; Dauphin & Vilmer 2007), while the GOES hard band peaked later, at 00:35. The sudden hard X-ray increase corresponds to the onset of complex type III emission, CME liftoff, a strong supra-arcade downflow seen with TRACE (Asai et al. 2004), and the peak in the derivative of the GOES X-ray flux (Reiner et al. 2007), indicating a strong injection of energetic electrons. The time of this major injection corresponds to the peak of the O VI emission obtained with UVCS, which peaked during the exposure between 00:26 and 00:28 UT. The first two UVCS detections during the rise phase correspond to the coronal hard X-ray source, while footpoint emission began at 00:27 (Krucker et al. 2003).

The SEE experiment detected this flare in the UV, EUV and X-rays (Woods et al., 2005), and Share et al. (2004) analyzed RHESSI observations of the positron annihilation line. Caspi & Lin (2010) find evidence for a superhot component to the X-ray emission in addition to the roughly 20 MK hot plasma. The temperatures in Table 2 derived from the GOES observations are similar to the hot plasma temperatures during the first 3 UVCS exposures and closer to the superhot temperatures during the last two. Fivian et al. (2009) derived energy deposition rates from RHESSI spectra of about $2.5 \times 10^{28}$ and $1.3 \times 10^{28}$ ergs s$^{-1}$ for times corresponding to the UVCS exposures beginning at 00:20 and 00:30 UT. Those values are about 10 times the luminosity of the X-ray emitting gas and about 60 times the upper transition region luminosity. Mancuso & Avetta (2008) detect broad O VI profiles associated with the coronal shock in the UVCS exposures beginning at 00:30:32 UT. This is consistent with the rapid dimming at that time seen in Figure 3.

*2002 Aug 24:* UVCS observations of this event are discussed by (Raymond et al. 2003). The peak of the O VI emission coincides with the peaks in microwave and hard X-ray emission (Reznikova et al. 2009), but precedes the GOES hard X-ray peak by about 12 minutes. Gamma ray emission detected by SONG began at 00:58 UT, near the time of the O VI peak, and continued for 9 minutes (Kuznetsov et al. 2006). Romano et al. (2009) studied the evolution of this flare. They found that the prominence became unstable at 00:52 UT, the same time that the RHESSI hard X-rays began to brighten. They also estimate densities of about $10^{11}$ cm$^{-3}$ during the time of the UVCS observations from the X-ray emission measures and TRACE loop sizes, rising to $3.5 \times 10^{11}$ at 01:10 UT. The Hard X-rays peaked at 00:56, about 4 minutes before the peak in O VI luminosity.

*2003 Nov 02:* The O VI emission from this event was also analyzed by Raymond et al. (2007). Woods & Rottman (2005) show that the soft X-ray emission, $\lambda < 40$Å, measured by XPS peaked 4 minutes later than the GOES emission. Share et al. (2004) show that the positron annihilation emission peaked at 17:20 UT, shortly after the O VI peak.

*2003 Nov 04:* The O VI emission from this event was also analyzed by Raymond et al. (2007), and UV emission from the current sheet was discussed by Ciaravella & Raymond (2008). The event showed two strong peaks in O VI, the one at 19:33 corresponding to spikes in the RHESSI count rate, and the brighter one at 19:42 corresponding to the sharp rise in GOES X-ray fluxes and saturation of the GOES detectors (Raymond et al. 2007, Fig. 1). The O VI emission is cut off when the CME crosses the UVCS slit position, so the second peak could occur later. Woods & Rottman (2005) show that the XPS soft X-rays peak 2 minutes after the GOES X-ray peak. Watanabe et al. (2006) find that the neutron event peaks at about 19:45, about 3 minutes after the UV peak. The SONG instrument on CORONAS-F reported strong gamma ray emission between 19:41 and 19:47 UT, with weaker emission continuing until 19:57 UT (Kuznetsov et al. 2006).

*2005 Jul 9:* This event was observed only in the LYA channel of UVCS, so no O VI measurements are available.

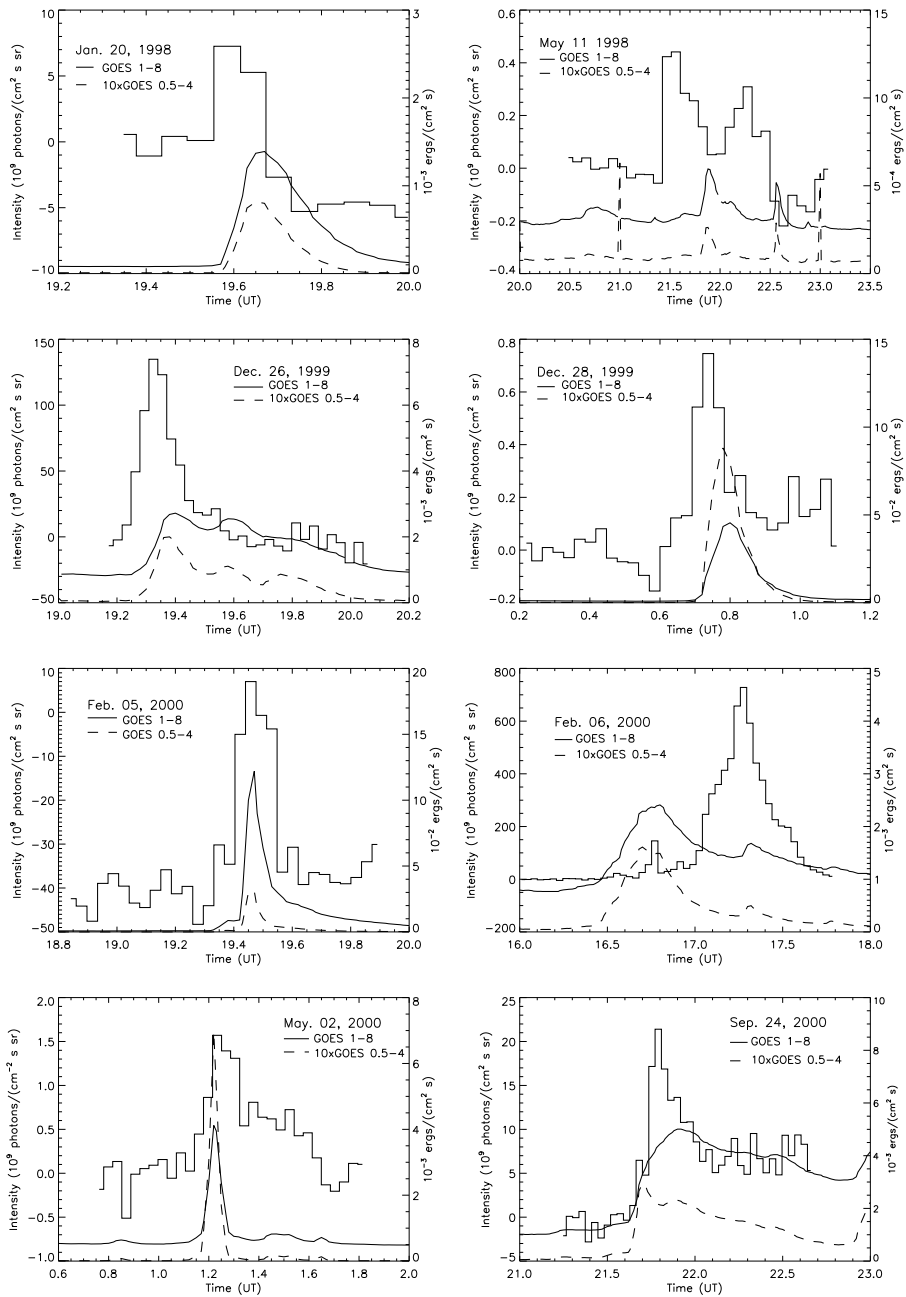

FIG. 1.— O VI λ1032 intensity after subtraction of the pre-event intensity (histogram) plotted against time for events through 2000 Sept. 24. The GOES soft X-ray flux is plotted as a solid line. The GOES hard X-ray flux is shown with the dashed line. The left-hand scale applies to O VI and the right-hand scale to GOES.

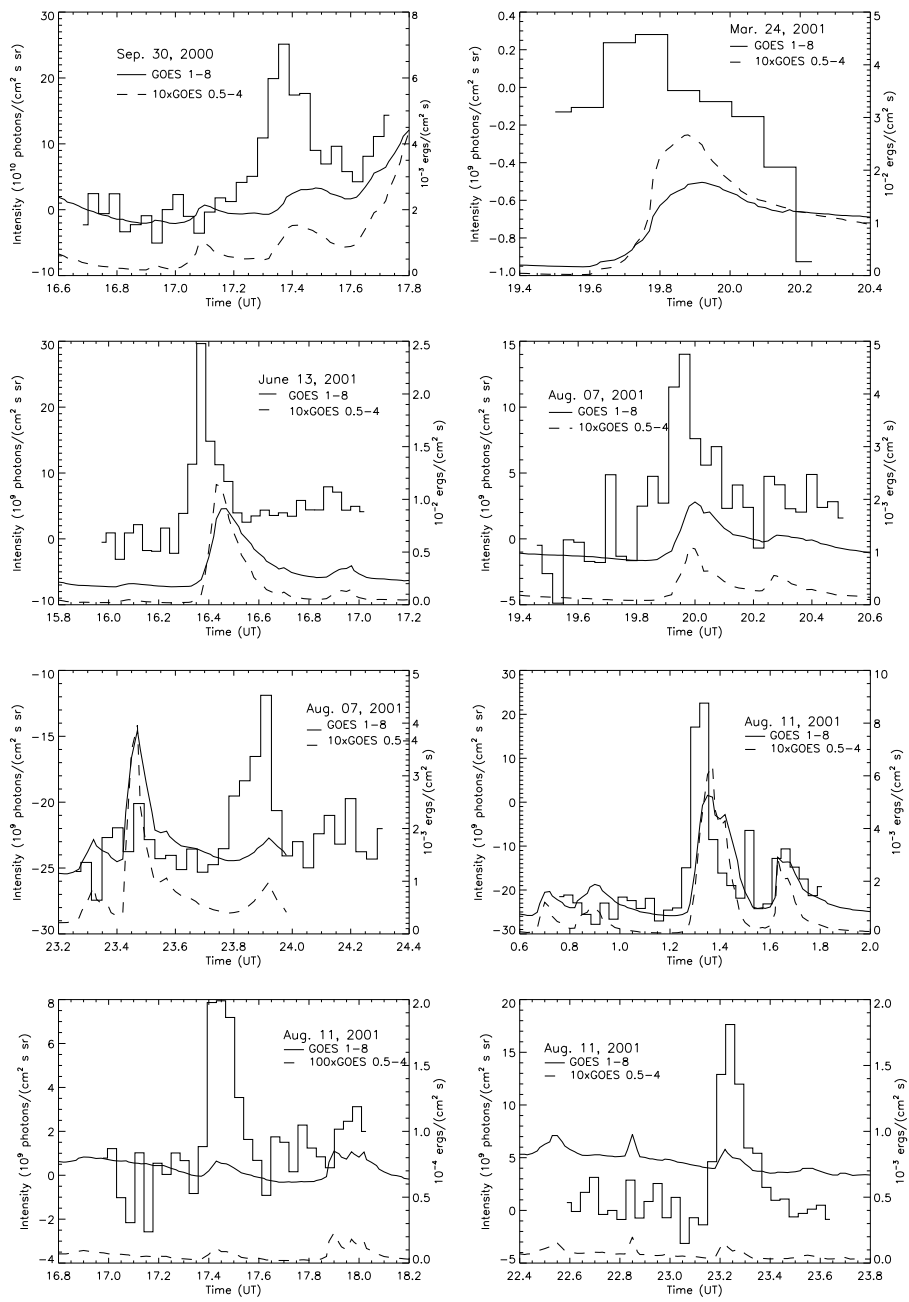

FIG. 2.— Same as Fig. 1 for events from 2000 Sept. 30 through 2001 Aug. 11.

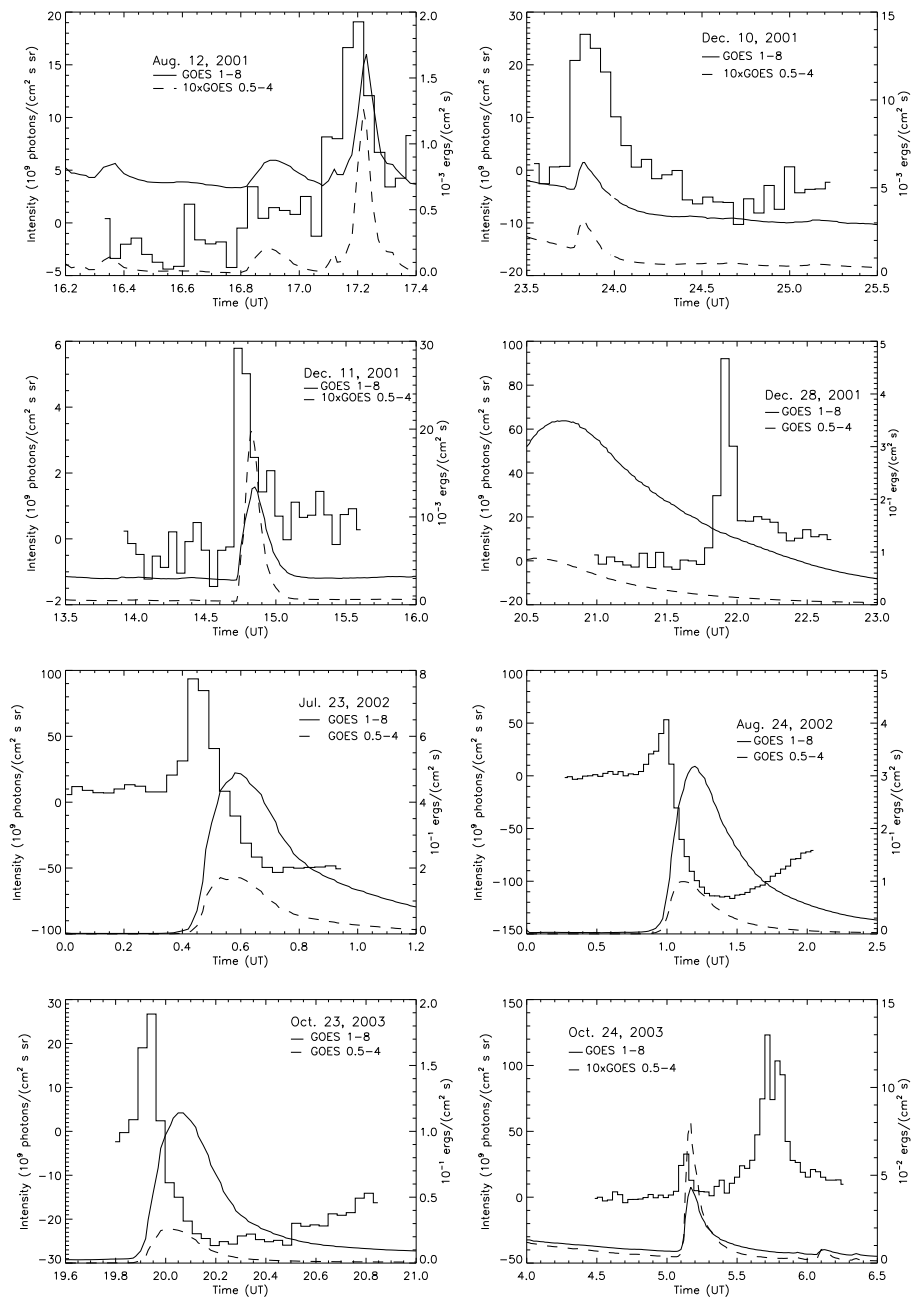

FIG. 3.— Same as Fig. 1 for events from 2001 Aug 12 through 2003 Oct. 24.

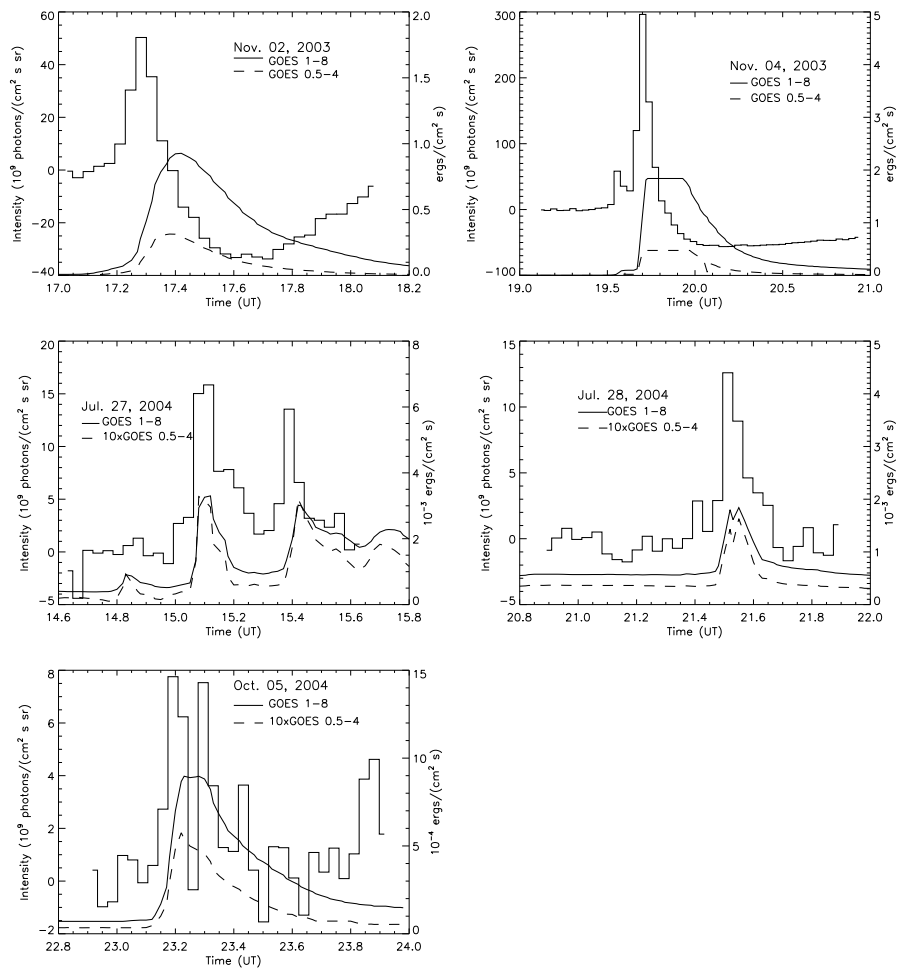

F<span style="font-variant:small-caps">ig</span>. 4.— Same as Fig. 1 for events from 2003 Nov. 2 through 2004 Oct. 5.

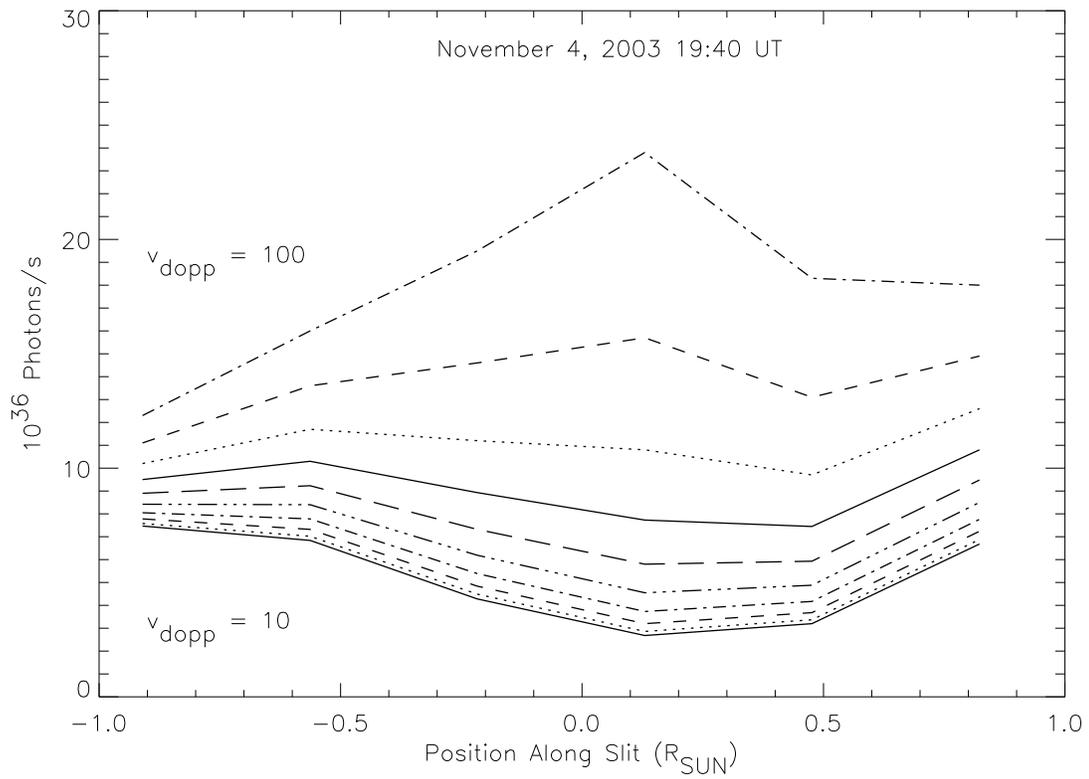

Fig. 5.— Apparent luminosity along the slit for flow speeds between 10 and 100 km s$^{-1}$ for the 2003 Nov. 4 event.

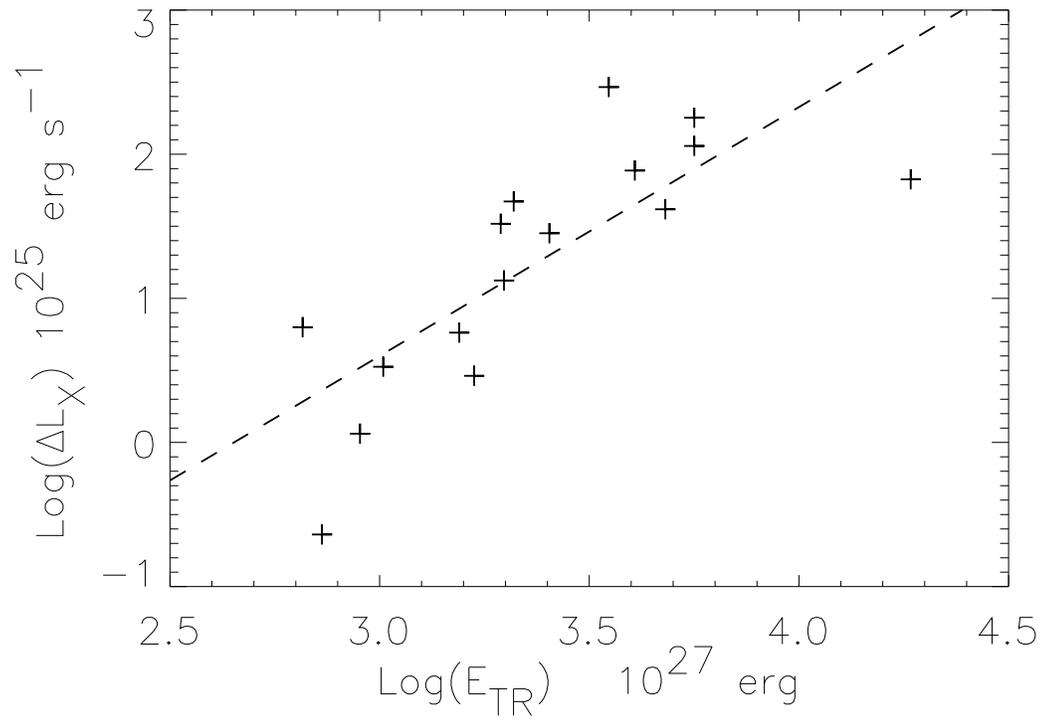

Fig. 6.— Transition region emission during the second UVCS exposure plotted against the change in GOES X-ray flux between the second and third UVCS exposures.

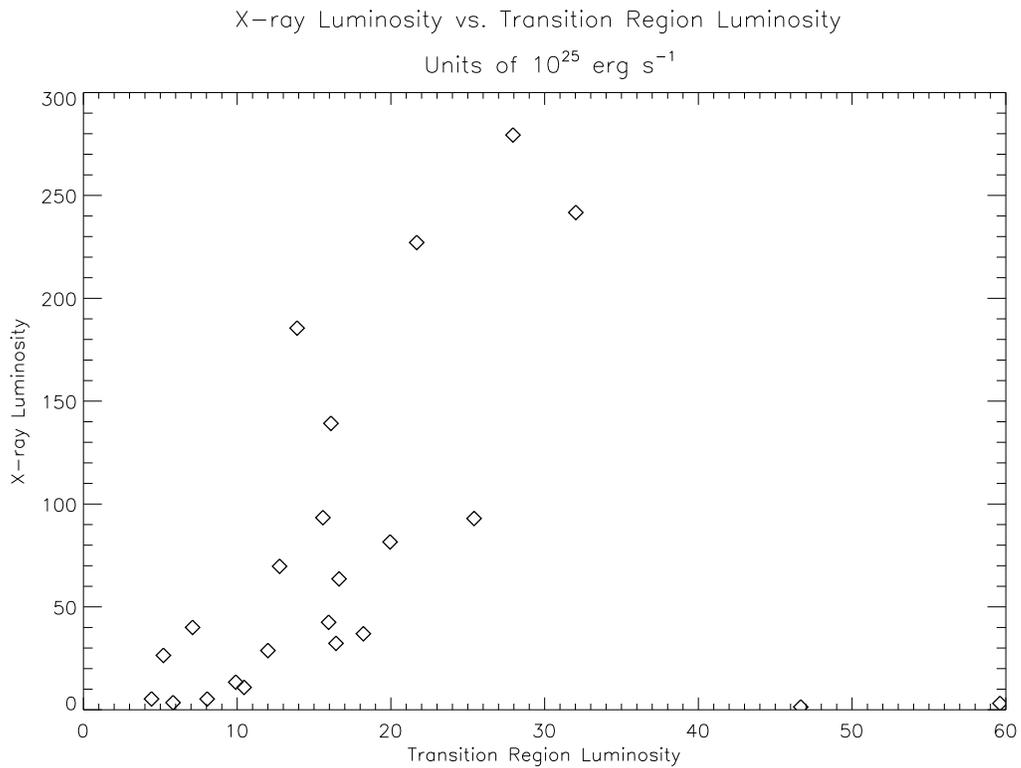

Fig. 7.— Transition region flux during the second UVCS exposure plotted against the GOES X-ray flux at the peak of the O VI luminosity.

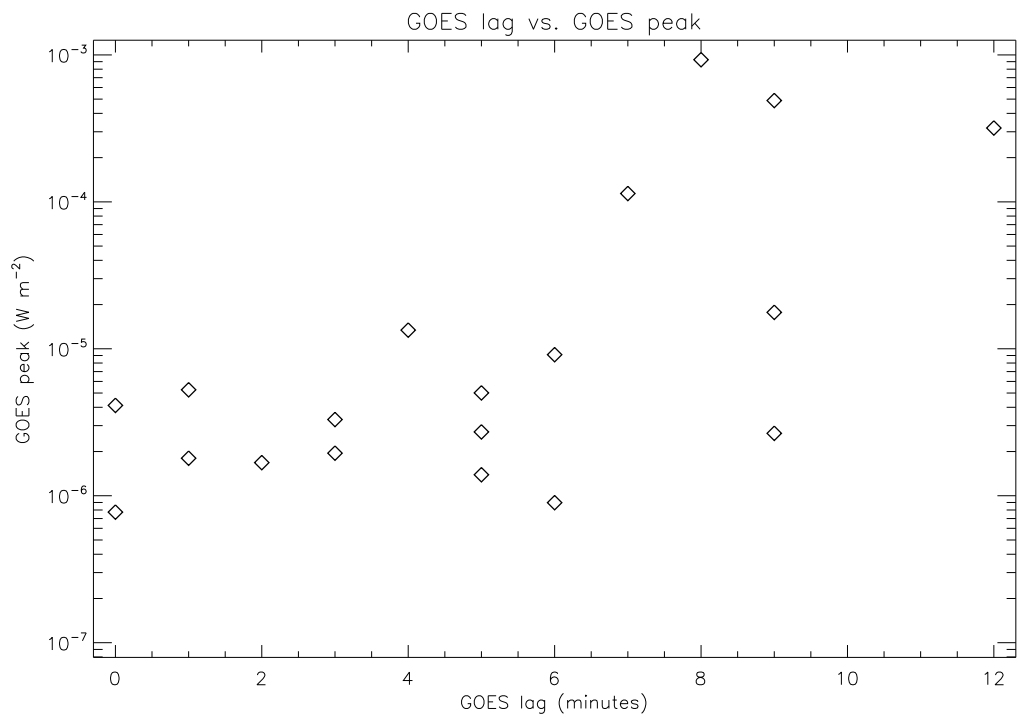

Fig. 8.— The peak flux in GOES soft X-rays plotted against the time lag between the peak of the O VI luminosity and the peak of the X-ray luminosity.

Table 1

UVCS Solar Flare Observations

| Event | Flare Class | Flare x,y ″,″ | Position Angle deg CCW | Height R$_{SUN}$ | Exposure sec | Slit $\mu m$ |
|---|---|---|---|---|---|---|
| SOL1998-01-20T19:39 | C1.4 | -891,559 | 60 | 1.67 | 200 | 50 |
| SOL1998-05-11T21:53 | B6.0[a] | -,- | 180 | 1.91 | 300 | 138 |
| SOL1999-12-26T19:24 | C2.7[a] | 978,-28 | 249 | 1.49 | 100 | 97 |
| SOL1999-12-28T00:43 | M5.6 | 673,545 | 249 | 2.52 | 100 | 97 |
| SOL2000-02-05T19:28 | X1.2[b] | -700,300 | 102 | 1.67 | 120 | 49 |
| SOL2000-02-06T16:48 | C2.4[b] | -930,-160 | 102 | 1.67 | 120 | 49 |
| SOL2000-05-02T01:13 | C4.1 | 865,-172 | 225 | 2.34 | 120 | 50 |
| SOL2000-09-24T21:54 | C5.0 | -944,170 | 110 | 1.64 | 120 | 97 |
| SOL2000-09-30T17:29 | C2.7 | -859,-507 | 100 | 1.63 | 120 | 97 |
| SOL2001-03-24T19:55 | M1.8 | -336,354 | 16 | 2.12 | 300 | 75 |
| SOL2001-06-13T16:28 | C9.1 | 680,426 | 263 | 1.62 | 120 | 100 |
| SOL2001-08-07T20:00 | C1.9 | 877,-239 | 245 | 1.63 | 120 | 96 |
| SOL2001-08-07T23:55 | C1.8 | 900,-350 | 245 | 1.63 | 120 | 96 |
| SOL2001-08-11T01:22 | C5.3 | 883,432 | 290 | 1.61 | 120 | 96 |
| SOL2001-08-11T17:26 | B7.7[a] | 920,300 | 290 | 1.61 | 120 | 96 |
| SOL2001-08-11T23:15 | B8.6 | 864,361 | 290 | 1.61 | 120 | 96 |
| SOL2001-08-12T17:14 | C1.7 | 862,545 | 290 | 1.61 | 120 | 97 |
| SOL2001-12-10T23:50 | C6.4 | 940,-198 | 263 | 1.56 | 200 | 96 |
| SOL2001-12-11T14:51 | M1.3 | 497,-286 | 263 | 1.91 | 200 | 96 |
| SOL2001-12-28 | ... | 979,277 | 282 | 1.54 | 200 | 97 |
| SOL2002-07-23T00:35 | X4.8 | -922,-213 | 96 | 1.63 | 120 | 97 |
| SOL2002-08-24T01:20 | X3.1 | 958,-33 | 260 | 1.63 | 120 | 97 |
| SOL2003-10-23T20:04 | X1.1 | -920,-295 | 106 | 1.69 | 120 | 96 |
| SOL2003-10-24T05:11 | M4.3 | -883,-245 | 106 | 1.69 | 120 | 96 |
| SOL2003-11-02T17:25 | X8.3 | 804,-200 | 245 | 1.63 | 120 | 97 |
| SOL2003-11-04T19:44 | >X28 | 963,-331 | 262 | 1.63 | 120 | 97 |
| SOL2004-07-27T15:07 | C3.3 | 807,20 | 270 | 1.77 | 120 | 97 |
| SOL2004-07-28T21:31 | C1.8 | 921,79 | 270 | 1.77 | 120 | 97 |
| SOL2004-10-05T23:14 | B9.0 | -942,217 | 75 | 1.72 | 120 | 97 |
| SOL2005-07-09T22:06 | M2.8 | 455,200 | 284 | 2.47 | 120 | 76 |

[a] Event behind limb
[b] EIT positions not available

Table 2

Impulsive Phase O VI and X-ray Parameters

Luminosities $10^{25}$ erg s$^{-1}$, Emission Measures $10^{48}$ cm$^{-3}$

| Date | Time UT | $L_{1032}$ $10^{25}$ | $L_{TR}$ $10^{25}$ | $L_X$ $10^{25}$ | $EM_{UV}$ $10^{48}$ | $EM_X$ $10^{48}$ | $T_6$ MK | $V_{dopp}$ km s$^{-1}$ |
|---|---|---|---|---|---|---|---|---|
| 1998 Jan 20 | 19:35:11 | 0.46 | 6.60 | 2.23 | 0.065 | 0.25 | 10.2 | |
| | 19:38:39 | 0.31 | 4.49 | 6.22 | 0.044 | 0.69 | 9.96 | |
| | 19:42:07 | 0.15 | 2.20 | 7.37 | 0.022 | 0.70 | 8.86 | |
| 1998 May 11 * | 21:28:04 | 5.13 | 73.9 | 0.548 | 0.723 | 0.05 | 5.24 | |
| | 21:33:35 | 4.65 | 67.0 | 0.869 | 0.656 | 0.08 | 5.83 | |
| | 21:38:48 | 3.83 | 55.1 | 1.28 | 0.539 | 0.11 | 5.71 | |
| | 21:44:22 | 4.12 | 59.3 | 1.26 | 0.580 | 0.12 | 5.30 | |
| | 21:49:33 | 0.61 | 8.82 | 1.34 | 0.086 | 0.13 | 9.14 | |
| | 21:55:04 | 1.10 | 15.8 | 2.80 | 0.155 | 0.23 | 7.71 | |
| 1999 Dec 26 * | 19:15:47 | 2.60 | 37.4 | 0.198 | 0.365 | 0.03 | 12.12 | 60 |
| | 19:17:38 | 3.94 | 56.8 | 0.749 | 0.555 | 0.12 | 12.25 | |
| | 19:19:29 | 5.96 | 85.9 | 1.68 | 0.840 | 0.31 | 12.93 | |
| | 19:21:17 | 4.52 | 65.1 | 3.56 | 0.637 | 0.65 | 12.93 | |
| | 19:23:08 | 4.32 | 62.2 | 5.69 | 0.608 | 0.85 | 12.08 | |
| | 19:24:59 | 3.52 | 50.7 | 6.64 | 0.496 | 0.90 | 11.06 | |
| 1999 Dec 28 | 0:42:24 | 2.37 | 34.1 | 26.7 | 0.334 | 7.76 | 19.6 | |
| | 0:44:15 | 2.45 | 35.2 | 203. | 0.344 | 62.1 | 20.2 | |
| | 0:46:06 | 1.86 | 26.8 | 495. | 0.262 | 135. | 17.6 | |
| 2000 Feb 05 | 19:25:19 | 1.90 | 27.3 | 26.7 | 0.267 | 6.86 | 16.7 | |
| | 19:27:28 | 2.07 | 29.8 | 116. | 0.291 | 33.7 | 18.9 | |
| | 19:29:39 | 1.88 | 27.1 | 102.6 | 0.265 | 20.6 | 14.6 | |
| | 19:31:48 | 2.15 | 30.9 | 63.2 | 0.303 | 12.7 | 13.9 | |
| 2000 Feb 06 | 16:31:00 | 0.97 | 13.9 | 14.9 | 0.136 | 2.23 | 11.9 | |
| | 16:33:13 | 1.28 | 18.5 | 88.1 | 0.181 | 14.6 | 12.3 | |
| | 16:35:25 | 1.47 | 21.2 | 123. | 0.208 | 20.4 | 12.5 | |
| | 16:37:34 | 0.90 | 12.9 | 28.2 | 0.126 | 4.67 | 12.3 | |
| 2000 May 02 | 1:09:37 | 0.51 | 7.35 | 11.4 | 0.072 | 2.09 | 13.2 | |
| | 1:11:49 | 1.14 | 16.4 | 46.9 | 0.161 | 10.31 | 15.6 | |
| | 1:13:58 | 1.70 | 24.5 | 60.2 | 0.239 | 11.0 | 13.4 | |
| | 1:16:07 | 1.32 | 19.1 | 43.4 | 0.186 | 4.10 | 8.66 | |
| | 1:18:18 | 1.65 | 23.7 | 22.7 | 0.232 | 1.90 | 7.75 | |
| 2000 Sep 24 | 21:43:06 | 0.08 | 1.12 | 81.2 | 0.011 | 12.1 | 12.0 | |
| | 21:45:15 | 1.21 | 17.4 | 129. | 0.171 | 15.7 | 10.6 | |
| | 21:47:24 | 1.35 | 19.5 | 176. | 0.190 | 19.5 | 9.74 | |
| | 21:49:35 | 1.24 | 17.8 | 214. | 0.175 | 21.8 | 9.53 | |
| | 21:51:45 | 1.05 | 15.1 | 226. | 0.147 | 23.0 | 9.49 | |
| | 21:53:55 | 1.16 | 16.7 | 235. | 0.163 | 23.9 | 9.41 | |
| | 21:56:08 | 0.67 | 9.68 | 237. | 0.095 | 24.1 | 9.27 | |
| 2000 Sep 30 | 17:17:57 | 0.42 | 6.01 | 6.35 | 0.059 | 1.16 | 13.1 | |
| | 17:20:06 | 0.89 | 12.8 | 6.52 | 0.125 | 1.43 | 15.3 | |
| | 17:22:17 | 0.74 | 10.6 | 12.3 | 0.104 | 2.71 | 14.8 | |
| | 17:24:27 | 0.68 | 9.83 | 18.8 | 0.096 | 3.78 | 14.1 | |
| | 17:26:36 | 0.71 | 10.2 | 23.1 | 0.100 | 4.22 | 13.4 | |
| 2001 Mar 24 | 19:46:33 | 1.80 | 25.9 | 184. | 0.253 | 40.4 | 15.6 | |
| | 19:52:03 | 1.21 | 17.5 | 270. | 0.171 | 59.5 | 15.3 | |
| 2001 Jun 13 | 16:20:01 | 1.01 | 14.5 | 4.80 | 0.142 | 0.88 | 12.8 | |
| | 16:22:10 | 2.50 | 36.0 | 23.3 | 0.352 | 4.26 | 13.3 | |
| | 16:24:22 | 1.36 | 19.6 | 64.7 | 0.192 | 14.2 | 15.6 | |
| | 16:26:31 | 0.92 | 13.2 | 139. | 0.129 | 28.0 | 14.2 | |
| | 16:28:40 | 1.14 | 16.4 | 175. | 0.160 | 29.1 | 12.5 | |
| 2001 Aug 07 | 19:55:41 | 0.64 | 9.15 | 16.2 | 0.090 | 2.18 | 11.3 | |
| | 19:57:51 | 1.17 | 16.9 | 30.5 | 0.165 | 4.56 | 11.6 | |
| | 20:00:04 | 0.69 | 9.99 | 39.7 | 0.098 | 5.35 | 11.0 | |
| 2001 Aug 07 | 23:48:05 | 0.34 | 4.90 | 14.4 | 0.048 | 1.20 | 7.46 | |
| | 23:50:15 | 0.56 | 8.06 | 13.1 | 0.079 | 1.12 | 7.91 | |
| | 23:52:25 | 0.39 | 5.62 | 11.3 | 0.055 | 1.06 | 8.74 | |
| | 23:54:38 | 0.56 | 8.06 | 10.7 | 0.079 | 1.09 | 9.20 | |

| Date | Time | | | | | | | |
|------|------|------|------|------|------|------|------|------|
| | 23:56:49 | 0.15 | 2.16 | 14.7 | 0.021 | 1.26 | 7.88 | |
| 2001 Aug 11 | 1:18:03 | 2.02 | 29.0 | 70.5 | 0.284 | 11.7 | 12.3 | 40 |
| | 1:20:12 | 2.18 | 31.4 | 114. | 0.307 | 18.8 | 13.0 | |
| | 1:22:23 | 1.10 | 15.8 | 94.8 | 0.155 | 17.36 | 13.5 | |
| * | 17:24:43 | 0.36 | 5.18 | 3.79 | 0.051 | 0.51 | 10.9 | |
| | 17:26:56 | 0.62 | 8.93 | 4.99 | 0.087 | 0.55 | 9.78 | |
| | 17:29:07 | 0.39 | 5.62 | 3.40 | 0.055 | 0.38 | 9.91 | |
| | 17:31:17 | 0.25 | 3.60 | 1.70 | 0.035 | 0.19 | 9.78 | |
| | 23:10:07 | 0.39 | 5.56 | 3.55 | 0.054 | 0.30 | 7.95 | |
| | 23:12:16 | 0.42 | 6.07 | 6.41 | 0.059 | 0.78 | 10.4 | |
| | 23:14:25 | 0.66 | 9.56 | 6.64 | 0.093 | 0.74 | 9.70 | |
| | 23:16:37 | 0.48 | 6.84 | 4.28 | 0.067 | 0.47 | 10.2 | |
| 2001 Aug 12 | 17:05:41 | 0.49 | 7.06 | 2.52 | 0.069 | 0.42 | 12.7 | 50 |
| | 17:07:50 | 0.59 | 8.50 | 3.61 | 0.083 | 0.49 | 11.2 | |
| | 17:10:01 | 0.85 | 12.2 | 6.96 | 0.120 | 1.04 | 12.0 | |
| | 17:12:10 | 0.96 | 13.8 | 15.1 | 0.135 | 3.05 | 13.7 | |
| | 17:14:19 | 0.74 | 10.7 | 26.3 | 0.105 | 3.93 | 11.9 | |
| 2001 Dec 10 | 23:46:31 | 0.32 | 4.62 | 25.8 | 0.045 | 4.28 | 12.3 | |
| | 23:50:00 | 0.41 | 5.84 | 41.9 | 0.057 | 6.26 | 11.9 | |
| | 23:53:30 | 0.40 | 5.82 | 30.6 | 0.057 | 4.12 | 10.9 | |
| | 23:57:00 | 0.31 | 4.51 | 7.46 | 0.044 | 0.70 | 8.78 | |
| 2001 Dec 11 | 14:43:48 | 1.29 | 18.6 | 74.3 | 0.182 | 13.6 | 13.6 | |
| | 14:47:18 | 1.41 | 20.3 | 133. | 0.199 | 31.7 | 15.9 | |
| | 14:50:48 | 0.65 | 9.42 | 210. | 0.092 | 38.5 | 13.6 | |
| 2001 Dec 28 | 21:51:20 | 1.44 | 20.8 | ... | 0.203 | ... | | 80 |
| | 21:54:50 | 4.64 | 66.8 | ... | 0.653 | ... | | |
| | 21:58:20 | 2.61 | 37.6 | ... | 0.368 | ... | | |
| 2002 Jul 23 | 0:21:51 | 0.67 | 9.66 | 7.66 | 0.094 | 1.68 | 15.2 | |
| | 0:24:02 | 0.88 | 12.6 | 14.6 | 0.124 | 4.00 | 18.3 | |
| | 0:26:12 | 3.29 | 47.4 | 42.9 | 0.463 | 14.3 | 22.9 | |
| | 0:28:21 | 2.73 | 39.3 | 133. | 0.384 | 49.5 | 29.4 | |
| | 0:30:32 | 1.42 | 20.4 | 256. | 0.199 | 92.5 | 27.3 | |
| 2002 Aug 24 | 0:51:03 | 0.62 | 8.90 | 12.8 | 0.087 | 2.12 | 12.6 | |
| | 0:53:15 | 0.97 | 14.0 | 12.6 | 0.137 | 2.77 | 15.4 | |
| | 0:55:24 | 1.92 | 27.6 | 15.5 | 0.270 | 4.25 | 18.0 | |
| | 0:57:33 | 3.15 | 45.4 | 28.3 | 0.444 | 9.41 | 23.0 | |
| | 0:59:44 | 4.37 | 63.0 | 63.3 | 0.616 | 23.2 | 28.3 | |
| | 1:01:54 | 2.22 | 31.9 | 112. | 0.312 | 41.4 | 28.8 | |
| 2003 Oct 23 | 19:54:27 | 0.94 | 13.6 | 21.1 | 0.133 | 6.75 | 21.6 | |
| | 19:56:37 | 1.12 | 16.2 | 52.6 | 0.158 | 17.6 | 21.7 | |
| | 19:58:46 | 0.71 | 10.3 | 89.5 | 0.101 | 27.4 | 20.1 | |
| | 20:00:57 | 0.77 | 11.0 | 116. | 0.108 | 33.6 | 18.6 | |
| 2003 Oct 24 | 5:06:05 | 1.78 | 25.6 | 55.0 | 0.250 | 13.1 | 15.8 | |
| | 5:08:17 | 3.26 | 46.9 | 302. | 0.459 | 82.8 | 17.2 | |
| | 5:10:26 | 0.79 | 11.4 | 481. | 0.111 | 123. | 16.5 | |
| 2003 Nov 2 | 17:12:37 | 1.15 | 16.6 | 72.9 | 0.162 | 23.3 | 21.2 | |
| | 17:14:48 | 3.26 | 46.9 | 126. | 0.459 | 41.8 | 22.5 | |
| | 17:16:58 | 4.84 | 69.7 | 240. | 0.682 | 86.6 | 26.3 | |
| | 17:19:07 | 2.02 | 29.1 | 419. | 0.285 | 153. | 27.2 | |
| 2003 Nov 4 | 19:31:12 | 1.56 | 22.5 | 23.8 | 0.220 | 3.20 | 11.5 | 80 |
| | 19:33:22 | 10.7 | 154. | 46.1 | 1.51 | 13.4 | 18.8 | |
| | 19:35:32 | 6.47 | 93.2 | 113. | 0.911 | 30.9 | 18.4 | |
| | 19:37:45 | 6.06 | 87.3 | 122. | 0.854 | 33.4 | 17.4 | |
| | 19:39:56 | 21.2 | 305. | 195. | 2.99 | 68.6 | 25.7 | |
| | 19:42:06 | 41.5 | 598. | ... | 5.84 | ... | ... | |
| | 19:44:15 | 12.3 | 177. | ... | 1.73 | ... | ... | |
| 2004 Jul 27 | 15:04:45 | 1.80 | 25.9 | 56.4 | 0.253 | 10.3 | 13.1 | |
| | 15:06:49 | 1.70 | 24.5 | 71.4 | 0.240 | 10.7 | 11.7 | |
| | 15:08:53 | 1.16 | 16.7 | 46.6 | 0.164 | 6.96 | 11.5 | |
| | 15:10:57 | 1.19 | 17.1 | 36.8 | 0.168 | 4.08 | 10.1 | |
| | 15:13:01 | 0.53 | 7.65 | 20.6 | 0.075 | 2.51 | 10.7 | |
| | 15:21:17 | 0.70 | 10.0 | 21.3 | 0.098 | 2.59 | 10.7 | |
| | 15:23:21 | 1.32 | 19.0 | 32.6 | 0.186 | 4.87 | 12.1 | |
| | 15:25:25 | 0.46 | 6.64 | 54.4 | 0.065 | 9.97 | 13.1 | |

| 2004 Jul 28 | 21:30:43 | 1.53 | 22.0 | 38.6 | 0.215 | 5.20 | 11.3 |
| | 21:32:47 | 1.25 | 17.9 | 32.7 | 0.176 | 4.90 | 12.1 |
| | 21:34:51 | 0.65 | 9.33 | 25.5 | 0.091 | 3.44 | 11.3 |
| 2004 Oct 05 | 23:11:30 | 1.06 | 15.20 | 28.9 | 0.149 | 3.52 | 10.5 |
| | 23:13:34 | 0.38 | 5.46 | 39.9 | 0.053 | 4.42 | 9.83 |
| | 23:15:38 | 0.48 | 6.90 | 46.1 | 0.067 | 4.69 | 9.42 |
| | 23:17:42 | 0.17 | 2.39 | 45.2 | 0.023 | 4.60 | 9.22 |

* Event Behind Limb

Table 3

Impulsive Phase Ly$\alpha$ (erg s$^{-1}$)

| Date | Time UT | $L_{Ly\alpha}$ $10^{25}$ |
|------|---------|------|
| 1998 Jan 20 | | |
| | 19:35:12 | 3.98 |
| | 19:38:40 | 2.32 |
| 1999 Dec 26 | | |
| | 19:19:30 | 11.5 |
| | 19:21:18 | 14.6 |
| | 19:23:09 | 11.2 |
| | 19:25:00 | 9.28 |
| | 19:26:48 | 6.61 |
| 2000 Feb 5 | | |
| | 19:25:20 | 6.57 |
| | 19:27:29 | 10.2 |
| | 19:29:40 | 9.20 |
| | 19:31:49 | 11.3 |
| 2000 Feb 6 | | |
| | 16:28:51 | 30.0 |
| | 16:31:01 | 39.3 |
| | 16:33:14 | 41.4 |
| | 16:35:26 | 50.5 |
| | 16:37:35 | 42.0 |
| 2005 Jul 09 | | |
| | 21:59:50 | 122. |
| | 22:01:54 | 162. |
| | 22:03:58 | 147. |
| | 22:06:02 | 145. |